\documentclass[journal]{IEEEtran}

\ifCLASSINFOpdf
\else
\usepackage[dvips]{graphicx}
\fi
\usepackage{url}

\usepackage{graphicx} 
\usepackage{booktabs} 
\usepackage{mathrsfs}
\usepackage{amsmath}
\usepackage{amssymb}
\usepackage{mathtools}
\usepackage{cite}
\usepackage[acronym,shortcuts]{glossaries}
\usepackage{lipsum}
\usepackage{xcolor}

\usepackage{comment}

\usepackage{algorithm, algpseudocode}

\usepackage{orcidlink}

\DeclareMathOperator*{\argmin}{arg\,min}

\newacronym{GA}{GA}{genie-aided}
\newacronym{EA}{EA}{``\emph{estimate-then-average}''}
\newacronym{AE}{AE}{``\emph{average-then-estimate}''}
\newacronym{IRS}{IRS}{intelligent reflecting surface}
\newacronym{RSSI}{RSSI}{received signal strength indicator}
\newacronym{SotA}{SotA}{state-of-the-art}
\newacronym{CSI}{CSI}{channel state information}
\newacronym{D2D}{D2D}{device-to-device}
\newacronym{RR}{RR}{round-robin}
\newacronym{DA}{DA}{Dutch auction}
\newacronym{CWFL}{CWFL}{clustered WFL}
\newacronym{WFL}{WFL}{wireless federated learning}
\newacronym{RSMA}{RSMA}{rate splitting multiple access}
\newacronym{IoT}{IoT}{Internet-of-Things}
\newacronym{TDMA}{TDMA}{time-domain multiple access}
\newacronym{NOMA}{NOMA}{non-orthogonal multiple access}
\newacronym{ML}{ML}{machine learning}
\newacronym{MIMO}{MIMO}{multiple-input multiple-output}
\newacronym{CT}{CT}{compute-then-transmit}
\newacronym{FP}{FP}{fractional programming}
\newacronym{CF-mMIMO}{CF-mMIMO}{cell free massive MIMO}
\newacronym{iid}{i.i.d.}{independent and identically distributed}
\newacronym{DL}{DL}{downlink}
\newacronym{UL}{UL}{uplink}
\newacronym{IC}{IC}{interference cancellation}
\newacronym{SIC}{SIC}{successive interference cancellation}
\newacronym{SI}{SI}{soft-impute}
\newacronym{BS}{BS}{base station}
\newacronym{TX}{TX}{transmit}
\newacronym{RX}{RX}{receive}
\newacronym{MU}{MU}{multi-user}
\newacronym{SISO}{SISO}{single-input single-output}
\newacronym{AWGN}{AWGN}{additive white Gaussian noise}
\newacronym{SINR}{SINR}{signal-to-interference-and-noise ratio}
\newacronym{FL}{FL}{federated learning}
\newacronym{CPU}{CPU}{central processing unit}
\newacronym{KNN}{KNN}{K-nearest-neighbor}
\newacronym{RF}{RF}{radio frequency}
\newacronym{GD}{GD}{gradient descent}

\newacronym{RSS}{RSS}{received signal strength}
\newacronym{FIM}{FIM}{fisher information matrix}
\newacronym{ToA}{ToA}{time of arrival}
\newacronym{AoA}{AoA}{angle of arrival}
\newacronym{GP}{GP}{Gaussian process}
\newacronym{2D}{2D}{two-dimensional}
\newacronym{GPR}{GPR}{Gaussian process regression}
\newacronym{GNSS}{GNSS}{global navigation satellite systems}
\newacronym{B5G}{B5G}{beyond fifth-generation}
\newacronym{RRH}{RRH}{remote radio head}
\newacronym{GPS}{GPS}{Global Positioning System}
\newacronym{RFID}{RFID}{radio frequency identification}
\newacronym{TCAS}{TCAS}{traffic alert and collision avoidance systems}
\newacronym{RMSE}{RMSE}{root mean square error}
\newacronym{SGD}{SGD}{stochastic gradient descent}
\newacronym{PDF}{PDF}{probability density function}
\newacronym{CU}{CU}{computing unit}
\newacronym{DM-MIMO}{DM-MIMO}{distributed massive multiple-input multiple-output}
\newacronym{LOS}{LOS}{line-of-sight}
\newacronym{NLOS}{NLOS}{non-line-of-sight}
\newacronym{ROI}{ROI}{region of interest}
\newacronym{AP}{AP}{access point}
\newacronym{TDOA}{TDOA}{time difference of arrival}
\newacronym{UE}{UE}{user equipment}
\newacronym{dB}{dB}{decibel}
\newacronym{RIS}{RIS}{reconfigurable intelligent surface}
\newacronym{CG}{CG}{conjugate gradient}

\newacronym{PG}{PG}{proximal gradient}
\newacronym{SVT}{SVT}{singular value thresholding}
\newacronym{NN}{NN}{nuclear norm}
\newacronym{NMSE}{NMSE}{normalized mean square error}
\newacronym{MC}{MC}{matrix completion}
\newacronym{NP}{NP}{non-deterministic polynomial-time}
\newacronym{SDP}{SDP}{semidefinite programming}

\begin{document}

\title{Discrete Aware Matrix Completion via\\ Convexized $\ell_0$-Norm Approximation}

\author{Niclas~F\"uhrling\textsuperscript{\orcidlink{0000-0003-1942-8691}}, Kengo Ando\textsuperscript{\orcidlink{0000-0003-0905-2109}}, \IEEEmembership{Graduate Student Members, IEEE},\\
Giuseppe~Thadeu~Freitas~de~Abreu\textsuperscript{\orcidlink{0000-0002-5018-8174}}, David~Gonz{\'a}lez~G.\textsuperscript{\orcidlink{0000-0003-2090-8481}}, \IEEEmembership{Senior Members, IEEE} and Osvaldo~Gonsa\textsuperscript{\orcidlink{0000-0001-5452-8159}}\\[-6ex]

\thanks{N.~F\"uhrling, Kengo Ando and G.~T.~F.~Abreu are with the School of Computer Science and Engineering, Constructor University, Campus Ring 1, 28759, Bremen, Germany (emails: [nfuehrling, kando, gabreu]@constructor.university).}
\thanks{D.~Gonz{\'a}lez~G. and O.~Gonsa are with the Wireless Communications Technologies Group, Continental AG, Wilhelm-Fay Strasse 30, 65936, Frankfurt am Main, Germany (e-mails: david.gonzalez.g@ieee.org, osvaldo.gonsa@continental-corporation.com).}
}

\maketitle

\begin{abstract}
We consider a novel algorithm, for the completion of partially observed low-rank matrices in a structured  setting where each entry can be chosen from a finite discrete alphabet set, such as in common recommender systems.
The proposed low-rank \ac{MC} method is an improved variation of \ac{SotA} discrete aware matrix completion method which we previously proposed, in which discreteness is enforced by an $\ell_0$-norm regularizer, not by replaced with the $\ell_1$-norm, but instead approximated by a continuous and differentiable function normalized via \ac{FP} under a \ac{PG} framework.
Simulation results demonstrate the superior performance of the new method compared to the \ac{SotA} techniques as well as the earlier $\ell_1$-norm-based discrete-aware matrix completion approach.

\end{abstract}


\IEEEpeerreviewmaketitle

\vspace{-3ex}
\section{Introduction}

\IEEEPARstart{W}{ith} the recent popularity of machine learning and big data, matrix completion has become a problem in many modern applications stemming from different areas, such as recommendation systems in computer science \cite{Chen_2022}, localization algorithms in signal processing \cite{Nguyen_2019_Loc}, and millimeter wave channel estimation in wireless communications \cite{Vlachos_2018}.

The most common problems are often structured low-rank \ac{MC} problems, where $\boldsymbol{X}\in \mathbb{R}^{m\times n}$ is the desired low-rank matrix obtained from the partially observed, incomplete matrix $\boldsymbol{O}\in \mathbb{R}^{m\times n}$ \cite{Nguyen_2019 , Dai_2012 , Bart_2013}, and traditional \ac{MC} solutions \cite{Candes_2009_Noise,Candes_2009,Candes_2010} mainly seek to improve performance and or lower complexity by replacing the non-convex rank objective with corresponding convex alternatives, \emph{i.e.} the \ac{NN}.

To cite a few \ac{SotA} examples, an \ac{MC} approach employing the \ac{NN} as a bound on the rank objective was proposed in \cite{Candes_2009_Noise}, which was shown to be able to fill matrices even with a few noisy observations.
Another approach proposed in \cite{Candes_2009} describes the relaxed minimization of the \ac{NN}, solved via \ac{SDP}.
Since the complexity of the \ac{SDP} solution is of order $\boldsymbol{O}(\text{max}(m,n)^4)$, where $m$ and $n$ are the dimensions of the matrix, the method can only solve the problem up to a certain size before becoming unfeasible.
In order to lower the complexity of the problem, the technique proposed in \cite{Cai_2010} relies on \ac{SVT} as a \ac{PG} minimizer of the \ac{NN} function.

Motivated by that idea, we recently proposed in \cite{Iimori_2020} a similar \ac{PG}-based low-complexity low-rank \ac{MC} method to solve a harder variation of the problem faced when the matrix entries must belong to a finite discrete alphabet.
The resulting discrete-aware \ac{MC} technique is especially efficient to solve problems such as in recommendation and rating applications \cite{Chen_2022}.
This discreteness in the previously proposed method is included by adding a discrete-space regularizer in the optimization problem, with the solution offered in closed-form under a relaxed variation in which the regularizer is convexized by using the $\ell_1$-norm.  

In this article, we propose an improvement of the latter discrete-aware matrix completion algorithm \cite{Iimori_2020} in which the discrete-space regularizer is the actual $\ell_0$-norm, relaxed by first replacing it with an arbitrarily-tight continuous and differentiable, but not convex, function, which is then convexized via \ac{FP}.
The method also makes use a \ac{PG} algorithm designed specifically for the formulated problem.

The structure of the article is as follows.
First, basics on the \ac{MC} problem, as well as a review of the discrete-aware variation of the \ac{MC} proposed in \cite{Iimori_2020} is given in Section \ref{sec:prior}.
Then, the new method based on the $\ell_0$-norm approximation is described in Section \ref{sec:ProposedMC}, followed by simulations results comparing our contribution with \ac{SotA} methods and a few concluding remarks, in Sections \ref{sec:results} and \ref{sec:conclusions}, respectively.

\section{Matrix Completion}
\label{sec:prior}

\subsection{Classical Formulation of the Matrix Completion Problem}
Before looking into the proposed method, a few major \ac{MC} techniques will be discussed, all of which aim to recover entries from a partially observed matrix.
Typical \ac{SotA} \ac{MC} techniques are based on a low-rank constraint, such that the original optimization problem can be written as
\vspace{-1ex}
\begin{subequations}
\label{eq:rank_opt}
\begin{align}
\argmin_{\boldsymbol{X}\in \mathbb{R}^{m\times n}}&\quad \text{rank}(\boldsymbol{X}),\\
\text{s.t. }&\quad P_{\Omega}(\boldsymbol{X})=P_{\Omega}(\boldsymbol{O}),
\vspace{-1ex}
\end{align}
\end{subequations}
where $\text{rank}(\cdot)$ indicates the rank of the input matrix, and $P_{\Omega}(\cdot)$ indicates a mask operator, defined as 
\vspace{-1ex}
\begin{equation}
[P_{\Omega}(\boldsymbol{X})]_{i,j}=
\begin{cases}
[\boldsymbol{X}]_{i,j},& \text{if } (i,j) \in \Omega, \\
0,              & \text{otherwise},
\end{cases}
\vspace{-1ex}
\end{equation}
where $\Omega$ denotes the index set of observed elements, and $[\cdot]_{i,j}$ denotes the $(i,j)$-th element of a given matrix.

We emphasize that although problem \eqref{eq:rank_opt} is optimal, it is also \ac{NP}-hard due to the non-convexity of the rank operator, such that its solution cannot be obtained efficiently.

It has also been shown \cite{Candes_2012}, however, that the problem can be relaxed replacing the rank objective by the \ac{NN} operator $||\boldsymbol{X}||_*$, which is given by the sum of the singular values of $\boldsymbol{X}$, since a matrix of rank $r$ has exactly $r$ nonzero singular values.
The relaxed optimization problem can be written as
\begin{subequations}
\label{eq:NNproblem}
\begin{align}
\argmin_{\boldsymbol{X}\in \mathbb{R}^{m\times n}}&\quad ||\boldsymbol{X}||_*,\\
\text{s.t. }&\quad P_{\Omega}(\boldsymbol{X})=P_{\Omega}(\boldsymbol{O}) ,
\label{eq:C1}
\end{align}
\end{subequations}
where the \ac{NN} can be viewed as a tight lower bound of the rank operator \cite{Recht_2010}, compared to the original problem in \eqref{eq:rank_opt}.


Taking into account realistic scenarios where observed matrices can contain noisy measurements, constraint \eqref{eq:C1} can be relaxed to the target matrix being recovered only as an approximately low-rank solution.
Various related works, $e.g.$ \cite{ OptSpace, RankEDM,Wong_2017}, have therefore considered the following variation of  problem \eqref{eq:NNproblem} 
%
%
\begin{subequations}
\label{eq:NN_opt}
\begin{align}
\argmin_{\boldsymbol{X}\in \mathbb{R}^{m\times n}}&\quad ||\boldsymbol{X}||_*,\\[-1ex]
\text{s.t. }&\quad \underbrace{\frac{1}{2}||P_{\Omega}(\boldsymbol{X}-\boldsymbol{O}) ||^2_F}_{\triangleq f(\boldsymbol{X})}\leq \epsilon,
\end{align}
\end{subequations}
where $f(\boldsymbol{X})$ was implicitly defined for notational convenience.

Problem \eqref{eq:NN_opt} can be rewritten in regularized form as
\begin{equation}
\label{eq:NN_opt_reg}
\argmin_{\boldsymbol{X}\in \mathbb{R}^{m\times n}}\quad f(\boldsymbol{X})+\lambda||\boldsymbol{X}||_*,
\end{equation}
or, including the rank constraint, as
\begin{subequations}
\label{eq:rank_opt2}
\begin{align}
\argmin_{\boldsymbol{X}\in \mathbb{R}^{m\times n}}&\quad f(\boldsymbol{X}),\\
\text{s.t. }&\quad \text{rank}({\boldsymbol{X}})\leq \nu.
\end{align}
\end{subequations}

While most \ac{SotA} methods \cite{Meka_2009,Hu_2013,Tanner_2015} seek to solve variations of the aforementioned convex problems, significant progress has been made recently on non-convex optimization methods \cite{Wang_2021,Sun_2016,Li_2018}, which typically lead to better performance over convex alternatives due to the tighter proximity to the original (typically non-convex) formulations.
As an example, an excellent survey on non-convex optimization methods applied to low-rank \ac{MC} can be found in \cite{Chi_2019}.

%

\subsection{\Acl{SI} Approach}
\Ac{SI} is a recently proposed method to solve large-scale \ac{MC} problems \cite{Recht_2013,Fang_2017,Yao_2019} similar to those described by equations \eqref{eq:NN_opt} and \eqref{eq:NN_opt_reg}. 
The standard \ac{SI} method consists of the recursion
\begin{equation}
\label{eq:soft_inpute}
\boldsymbol{X}_t=\text{SVT}_\lambda(\boldsymbol{X}_{t-1}+P_{\Omega}(\boldsymbol{O}-\boldsymbol{X}_{t-1})),
\end{equation}
where the SVT function is given as
\begin{equation}
\text{SVT}_\lambda(\boldsymbol{A})=\boldsymbol{U}(\boldsymbol{\Sigma}-\lambda\boldsymbol{I})_+\boldsymbol{V}^\intercal,
\end{equation}
with $\boldsymbol{A}\triangleq\boldsymbol{U}\boldsymbol{\Sigma}\boldsymbol{V}^\intercal$ and $(\cdot)_+$ being the positive part of the input.

It has been shown in \cite{Yao_2019} that \ac{SI} can be used as a \ac{PG} method, enabling the use of \ac{SI}-based Nesterov-type momentum acceleration \cite{Nesterov}, expressed by the updated recursion
\begin{equation}
\boldsymbol{X}_t=\text{SVT}_\lambda(\boldsymbol{Y}_t+P_{\Omega}(\boldsymbol{O}-\boldsymbol{Y}_t)),
\end{equation}
with $\boldsymbol{Y}_t=(1+\gamma_t)\boldsymbol{X}_{t-1}+\gamma_t\boldsymbol{X}_{t-2}$, where $\gamma_t$ acts as the weight of the momentum.

\subsection{Discrete-Aware Matrix Completion}

The discrete-aware matrix completion scheme first proposed in \cite{Iimori_2020} assumes that the missing entries of the matrix that need be recovered belong to a finite discrete alphabet set $\mathcal{A}\triangleq\{a_1,a_2,\cdots\}$.
Thus, the standard regularized optimization problem discussed above can be extended to
\begin{equation}
\argmin_{\boldsymbol{X}\in \mathbb{R}^{m\times n}}\quad f(\boldsymbol{X})+\lambda g(\boldsymbol{X})+\zeta r(\boldsymbol{X}|p),
\end{equation}
where $g(\boldsymbol{X})$ denotes a low-rank regularizer, which is chosen to be the \acf{NN}, while the discrete-space regularizer $r(\boldsymbol{X}|p)$ defined as
\begin{equation}
r(\boldsymbol{X}|p)\triangleq \sum_{k=1}^{|\mathcal{A}|}||\text{vec}_{\bar{\Omega}}(\boldsymbol{X})-a_k\boldsymbol{1}||_p,
\end{equation}
with $0\leq p$, $|\mathcal{A}|$ being the cardinality of the set and $\text{vec}_{\bar{\Omega}}(\boldsymbol{X})$ representing a vectorization of the matrix $\boldsymbol{X}$, where the entries are chosen corresponding to the given index set $\bar{\Omega}$, being the complementary set to the previously presented set $\Omega$.


Following \ac{SotA} methods, a \ac{PG} method was used in \cite{Iimori_2020} to solve the problem iteratively, containing the following steps:
\begin{eqnarray}
&\boldsymbol{Y}_t=(1+\gamma_t)\boldsymbol{X}_{t-1}+\gamma_t\boldsymbol{X}_{t-2},\label{eq:acc_old}&\\
&\boldsymbol{Z}_t=\text{prox}_{\zeta_r}(\boldsymbol{Y}_t),&\\
&\boldsymbol{X}_t=\text{SVT}_\lambda(P_{\bar{\Omega}}(\boldsymbol{Z}_t)+P_{\Omega}(\boldsymbol{O})),&
\end{eqnarray}
where the first step integrates the moment acceleration function, the second corresponds to the proximal operation on the discrete space regularizer, and the last amounts to the \ac{PG} operation on the \ac{NN} regularizer $g(\boldsymbol{X})$, since it is known \cite{Cai_2010} that \ac{SVT} can be used as a proximal minimizer for the \ac{NN}.

Following the approach described in \cite{antonello2020proximal}, the operation on the discrete-space regularizer $\text{prox}_{\zeta_r}(\boldsymbol{Y}_t)$ is given by
\begin{equation}
\text{prox}_{\zeta_r}(\boldsymbol{Y}_t)\triangleq \argmin_{\boldsymbol{U}}\; r(\boldsymbol{U}|1)\! +\!\frac{1}{2\zeta}||\text{vec}_{\bar{\Omega}}(\boldsymbol{U}\!-\!\boldsymbol{Y}_t)||_2^2,
\end{equation}
where it was shown in \cite{Iimori_2020} that by an element-by-element reformulation, a closed-form solution can be found for the desired matrix $\boldsymbol{U}$, leading to
\begin{equation}
\Bar{\boldsymbol{u}}=\text{sign}(\Bar{\boldsymbol{y}}_t)\odot (|\Bar{\boldsymbol{y}}_t|-\zeta\boldsymbol{1})_+,
\end{equation}
as a preliminary solution, where $\Bar{u}_l\triangleq [\boldsymbol{u}]_l-a_k$ and $\Bar{y}_{t,l} \triangleq [\boldsymbol{y}_t]_l-a_k$ are used as auxiliary variables.
Finally,  $\boldsymbol{u}$ can be recovered by 
\begin{equation}
\boldsymbol{u}=\Bar{\boldsymbol{u}}+a_k\boldsymbol{1},
\end{equation}
which is then the final solution for the first proximal operation.

\section{Proposed Discrete-Aware Matrix Completion}
\label{sec:ProposedMC}

\subsection{Discrete-Aware Regularizer with $\ell_0$-Norm Approximation}
\label{sec:norm_approx}

Consider the following tight approximation of the $\ell_0$-norm proposed in \cite{Mohimani_2009} 
\begin{equation}
||\boldsymbol{x}||_0\approx\sum^L_{i=1}\frac{|x_i|^2}{|x_i|^2+\alpha}=L-\sum^L_{i=1}\frac{\alpha}{|x_i|^2+\alpha},
\end{equation}
where $L$ and $\alpha$ are the length of the vector $\boldsymbol{x}$ and an arbitrary small constant respectively.

Notice in particular that under this approximation we have, for any value of $x_i$,
\vspace{-1ex}
\begin{eqnarray}
&\dfrac{|x_i|^2}{|x_i|^2+\alpha}\rightarrow 0 \text{ for } |x_i|^2=0,&\\
&\dfrac{|x_i|^2}{|x_i|^2+\alpha}\rightarrow 1 \text{ for } |x_i|^2>0.&
\end{eqnarray}

In order to apply the approximation to the discrete-aware regularizer, we first rewrite it in an element-by-element format, using the approximation to each term.
Proceeding as such, the regularizer can be written as
\vspace{-1ex}
\begin{align}
\label{eq:r_l0}
r(\boldsymbol{X}|0)&=\sum_{k=1}^{|\mathcal{A}|}||\text{vec}_{\bar{\Omega}}(\boldsymbol{X})-a_k\boldsymbol{1}||_0=\sum_{k=1}^{|\mathcal{A}|}\sum_{j=1}^{|\bar{\Omega}|}||\boldsymbol{x}_j-a_k||_0\nonumber \\ &\approx\sum_{k=1}^{|\mathcal{A}|}\bigl(\underbrace{|\bar{\Omega}|}_{const.}-\sum_{j=1}^{|\bar{\Omega}|}\frac{\alpha}{|\boldsymbol{x}_j-a_k|^2+\alpha}\bigr)\nonumber\\
&=-\sum_{k=1}^{|\mathcal{A}|}\sum_{j=1}^{|\bar{\Omega}|}\frac{\alpha}{|\boldsymbol{x}_j-a_k|^2+\alpha},
\end{align}
where $|\bar{\Omega}|$ denotes the number of elements in the set $\bar{\Omega}$.
Additionally, since $|\bar{\Omega}|$ is a constant term, it can be neglected in the following due to working on an optimization problem.

\subsection{Reformulation via Fractional Programming}

Notice that the discrete-aware regularizer formulation of equation \eqref{eq:r_l0} is not convex due to the fraction of affine and convex functions.
In order to mitigate this challenge, we introduce a fractional-programming variation of the $\ell_0$-norm approximation, enabled by the quadratic transform described in \cite{Shen_2018P1, Shen_2018P2}.


A simple reformulation can be found by applying the quadratic transform to the original $\ell_0$-norm approximation, which yields
\vspace{-1ex}
\begin{eqnarray}
\label{eq:l0_fp}
||x||_0 \hspace{-4ex}&& \approx\! L\!-\!\sum^L_{i=1}\frac{\alpha}{|x_i|^2\!+\!\alpha}\!\approx\!L\!-\!\bigg(\sum^L_{i=1}2\beta_i \sqrt{\alpha}\!-\!\beta_i^2(|x_i|^2\!+\!\alpha)\!\bigg)\nonumber\\[-1ex]
&&=\!\sum^L_{i=1}\!\beta_i^2|x_i|^2\!+\!\underbrace{L\!-\!\bigg(\!\sum^L_{i=1}2\beta_i \sqrt{\alpha}\!+\!\alpha\!\bigg)}_{\text{indep. of $x$}}\!\equiv\! \sum^L_{i=1}\beta_i^2|x_i|^2,
\end{eqnarray}
where $\beta_i\triangleq\frac{\sqrt{\alpha}}{|x_i|^2+\alpha}$ is an auxiliary variable resulting from the fractional programming step, and the terms independent of $x$ can be neglected for the purpose of the minimization problem, leading to the last equivalent.

In the following, the quadratic transform can be applied to the problem statement of the discrete-space regularizer similar as shown in \eqref{eq:l0_fp}, leading to
\begin{equation}
-\sum_{k=1}^{|\mathcal{A}|}\sum_{j=1}^{|\bar{\Omega}|}\frac{\alpha}{|\boldsymbol{x}_j-a_k|^2+\alpha}=\sum_{k=1}^{|\mathcal{A}|}\sum_{j=1}^{|\bar{\Omega}|}\beta_{k,j}^2|\boldsymbol{x}_j-a_k|^2, 
\end{equation}
with $\beta_{k,j}\triangleq\frac{\sqrt{\alpha}}{|\boldsymbol{x}_j-a_k|^2+\alpha}$ being an auxiliary variable resulting from the quadratic transform as stated above.

In order to simplify and vectorize the regularizer to its final form, the following auxiliary vector and matrix are introduced

\quad\\[-4.5ex]
\begin{subequations}
\label{eq:updateBb}
\begin{equation}
\boldsymbol{b}=\sum_{k=1}^{|\mathcal{A}|}a_k[\beta_{k,1}^2,\beta_{k,2}^2,\cdots \beta_{k,|\bar{\Omega}|}^2]^\intercal,
\end{equation}
\begin{equation}
\boldsymbol{B}=\sum_{k=1}^{|\mathcal{A}|}\text{diag}(\beta_{k,1}^2,\beta_{k,2}^2,\cdots \beta_{k,|\bar{\Omega}|}^2)\succ 0,
\end{equation}
\end{subequations}
such that after some algebra, the final regularizer can be reformulated as
\begin{equation}
\label{eq:convexregularizer}
r(\boldsymbol{X}|0)\!\approx\! h(\boldsymbol{X})\!=\! \text{vec}_{\bar{\Omega}}(\boldsymbol{X})^\intercal\! \boldsymbol{B}\text{vec}_{\bar{\Omega}}(\boldsymbol{X})\! -\!2\text{vec}_{\bar{\Omega}}(\boldsymbol{X})^\intercal \boldsymbol{b}.\!
\end{equation}


Having obtained a closed-form and convex formulation of the approximate $\ell_0$-norm regularizer, the proposed optimization problem can be written as
\begin{equation}
\argmin_{\boldsymbol{X}\in \mathbb{R}^{m\times n}} {\underbrace{f(\boldsymbol{X})}_{\mathclap{\frac{1}{2}||P_{\Omega}(\boldsymbol{X}-\boldsymbol{O}) ||^2_F}}}+\lambda {\overbrace{g(\boldsymbol{X})}^{\mathclap{||\boldsymbol{X}||_*}}}+\zeta \underbrace{r(\boldsymbol{X}|0)}_{\text{eq. \eqref{eq:convexregularizer}}},
\end{equation}
or, using equation \eqref{eq:convexregularizer}, as
\begin{equation}
\label{eq:finalproblem}
\argmin_{\boldsymbol{X}\in \mathbb{R}^{m\times n}} {f(\boldsymbol{X})+{\zeta h(\boldsymbol{X})}}+\lambda g(\boldsymbol{X}).
\end{equation}


Now, the proximal gradient algorithm corresponding to this problem can be obtained by modifying the one originally derived in \cite{Iimori_2020}.
To that end, consider first the gradient of $h(\boldsymbol{X})$ be expressed by
\begin{equation}
\label{eq:grad_h}
\nabla h(\boldsymbol{X})\!=\text{vec}_{\bar{\Omega}}^{-1}(2\boldsymbol{B}\text{vec}_{\bar{\Omega}}(\boldsymbol{X})\!-\!2\boldsymbol{b}).
\end{equation}

Then, let us compute the momentum acceleration function as in \eqref{eq:acc_old}, namely
\begin{equation}
\boldsymbol{X}_t=(1+\gamma_t)\boldsymbol{X}_{t-1}+\gamma_t\boldsymbol{X}_{t-2},
\label{eq:Acc_new}
\end{equation}
where the momentum weights $\gamma_t$ are obtained as in \cite{Nesterov}.

Next, we process the SVT step where, as shown in \cite{antonello2020proximal}, the input corresponds to the matrix $\boldsymbol{X}_t$ and the gradient $\nabla h(\boldsymbol{X})$, which yields
\begin{align}
\boldsymbol{X}_{t+1}&=\text{SVT}_\lambda(\boldsymbol{X}_{t}-\nabla f(\boldsymbol{X}_t)-\mu \zeta\nabla h(\boldsymbol{X}_t)) \nonumber \\
&=\text{SVT}_\lambda(P_\Omega(\boldsymbol{O})+P_{\bar{\Omega}}(\boldsymbol{X}_{t}-\mu \zeta \nabla h(\boldsymbol{X_t}))),
\label{eq:SVT_new}
\end{align}
where $\mu$ denotes the a stepsize optimized for convergence guarantees by employing the Lipschitz condition
\begin{subequations}
\begin{equation}
0\leq\mu\leq 1/L,
\label{eq:step}
\end{equation}
with L denoting the Lipschitz constant, computed as
\label{eq:Lip}
\begin{equation}
L =\max\mathrm{eig} (\boldsymbol{B}^\intercal \boldsymbol{B}) =\mathrm{max}(B_{i,i}^2, \forall i ) \label{eq:Lip_low_comp}
\end{equation}
\end{subequations}
where $\max\mathrm{eig} (\cdot)$ denotes the operator returning the maximum eigenvalue of the argument matrix, and \eqref{eq:Lip_low_comp} offers a low complexity alternative.

The detailed procedure to solve problem \eqref{eq:finalproblem} is summarized as a pseudo-code in Algorithm \ref{alg:l0}.
The algorithm works in two loops until the convergence of $\boldsymbol{B}$ in the outer loop and the convergence of the gradient algorithm in the inner loop.
In the outer loop, the auxiliary variables $\boldsymbol{B}$ and $\boldsymbol{b}$, as well as the stepsize $\mu$ are calculated; while in the inner loop the first step is constructing $\boldsymbol{X}_t$ from the Nesterov-type momentum acceleration method, followed by the gradient $ h(\boldsymbol{X})$, and finally the soft-input.


\begin{algorithm}[H]
\caption{Discrete-Aware Matrix Completion with $\ell_0$-Norm Approximation}
\begin{algorithmic}[1]
\small
\Statex \hspace{-4ex} \textbf{Input:} Incomplete matrix $\boldsymbol{O}$ and constants $\alpha, \lambda$ and $\zeta$.
\Statex \hspace{-4.4ex} \hrulefill \vspace{-0.3ex}
\State \textbf{Initialization:} Initialize random $\boldsymbol{X}_{t-1}=\boldsymbol{X}_{t-2}$
\For{until convergence of B}
\State Compute $\boldsymbol{B},\boldsymbol{b}$ via \eqref{eq:updateBb}
\State Find $\mu$ via \eqref{eq:Lip}
\For{until convergence of Gradient algorithm}
\State  Compute $\boldsymbol{X}_t$ via \eqref{eq:Acc_new}
\State Compute $\nabla h(\boldsymbol{X}_t)$ via \eqref{eq:grad_h}
\State Compute $\boldsymbol{X}_{t+1}$ via \eqref{eq:SVT_new}
\EndFor
\EndFor
\Statex  \hspace{-4ex}  \textbf{Output:} Completed matrix $\boldsymbol{X}$. 

\end{algorithmic}
\label{alg:l0}
\end{algorithm}

\section{Numerical Results}
\label{sec:results}

In this section we compare the proposed method to the \ac{SotA} techniques described in \cite{Mazumder_2010}, the accelerated and inexact \ac{SI} (AIS)-impute approach of \cite{Yao_2019}, and our own earlier discrete-aware method described in \cite{Iimori_2020}.
For comparison such comparisons, the MovieLens-100k data set\footnote{Dataset available at https://grouplens.org/datasets/movielens/} is chosen, which is commonly used in the \ac{MC} literature on recommendation methods, where each matrix entry has an integer rating varying from 1 to 5.
The chosen performance metric for comparison is the \ac{NMSE}, defined by
\begin{equation}
\text{NMSE}\triangleq\frac{||P_{\bar{\Omega}}(\boldsymbol{X}-\boldsymbol{O})||^2_F}{||P_{\bar{\Omega}}(\boldsymbol{O})||^2_F}.
\end{equation}

\begin{figure}[H]
\includegraphics[width=\columnwidth]{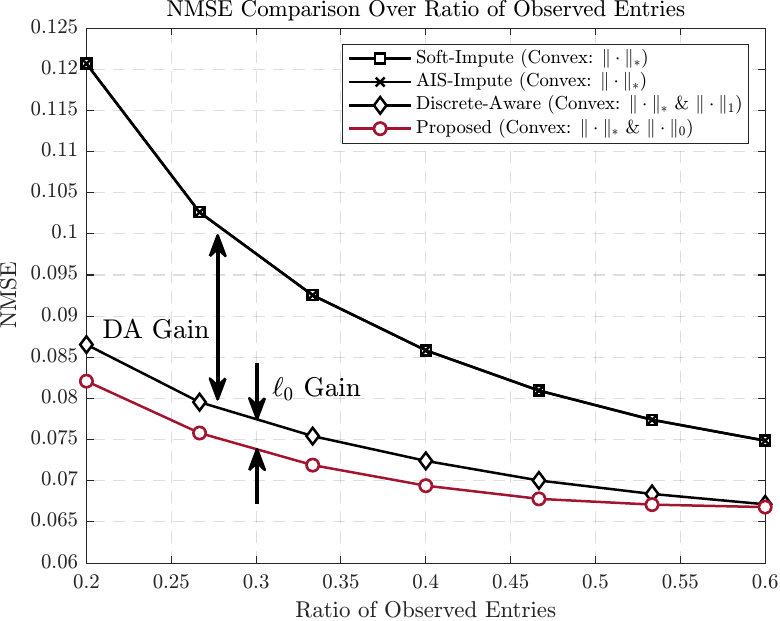}
\vspace{-4ex}
\caption{\ac{NMSE} comparison of the \ac{SotA} and the proposed method, with a varying ratio of observed entries in $\boldsymbol{O}$, for $\alpha=0.1$, $\lambda=10$ and $\zeta=0.15$.}
\label{fig:NMSE}
\end{figure}

\begin{figure}[H]
\includegraphics[width=\columnwidth]{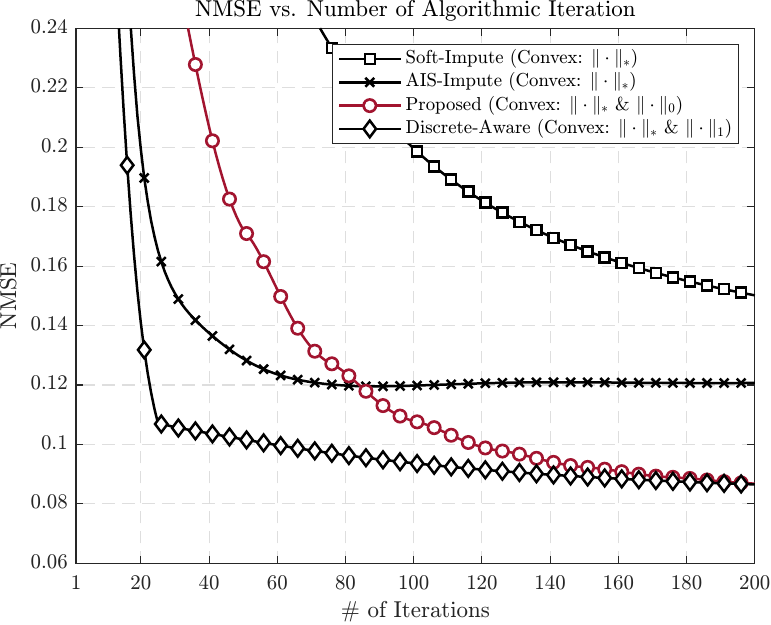}
\caption{\ac{NMSE} convergence comparison of the \ac{SotA} and the proposed method, with a $20\%$ observation ratio of $\boldsymbol{O}$, for $\alpha=0.1$, $\lambda=10$ and $\zeta=0.15$.}
\label{fig:conv}
\end{figure}

In order to submit both \ac{SotA} and proposed methods to the same conditions, identical simulation setups are chosen, where the ratio of observed values in the matrix $\boldsymbol{O}$ varies from $20\%$ to $60\%$.
The \ac{NMSE} results over varying observation ratios are shown in Figure \ref{fig:NMSE}, while the convergence behavior for a fixed observation ratio of $20\%$ is shown in Figure \ref{fig:conv}.

It is found from Figure \ref{fig:NMSE}, that indeed the proposed method outperforms all other alternatives, offering a gain over the earlier method of \cite{Iimori_2020} that is more significant at harsher conditions of where the observation ratios are lowest. 
\vspace{-1ex}

\begin{figure}[H]
\includegraphics[width=\columnwidth]{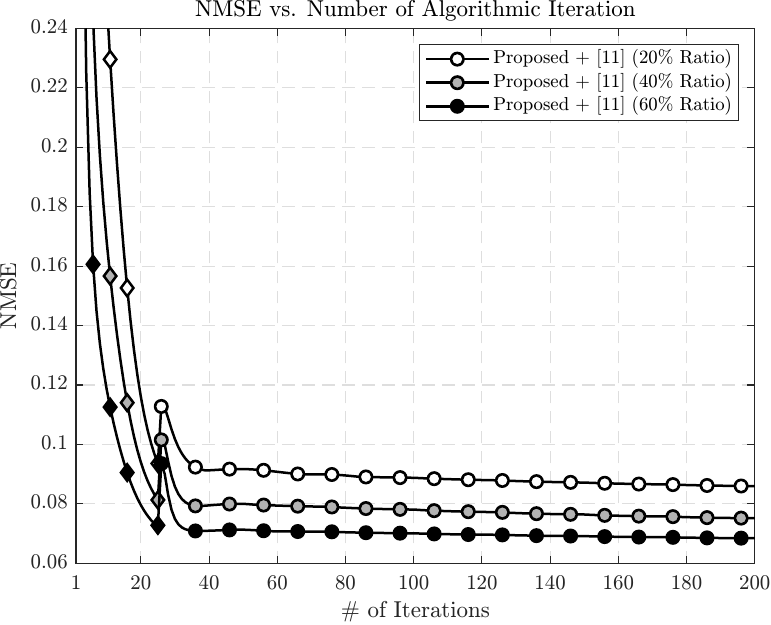}
\caption{\ac{NMSE} convergence comparison of the proposed method initialized with \cite{Iimori_2020}, with a varying ratio of observed entries in $\boldsymbol{O}$.}
\label{fig:conv_extd}
\end{figure}

In turn, our second set of results offered in Figure \ref{fig:conv} shows that the penalty of the gain achieved by the proposed method with asymptotically tight $\ell_0$-norm approximation, convexized via \ac{FP}, is a slower convergence compared to the earlier method based on the $\ell_1$-norm approximation of the $\ell_0$-norm, although the new method still converges as fast or faster than the techniques in \cite{Mazumder_2010} and \cite{Yao_2019}.
%
%
%
Given that the proposed method relies on a convexized problem, the convergence results of Figure \ref{fig:conv} suggests us to consider initializing the proposed method with the solution of our earlier technique from \cite{Iimori_2020}.
This results corresponding to this approach are shown in Figure \ref{fig:conv_extd}, and demonstrate that indeed, the combination of the earlier method from \cite{Iimori_2020} with the one here proposed yields the best combination of convergence and performance.

\section{Conclusion}
\label{sec:conclusions}

We proposes a new discrete-aware \ac{MC} algorithm in which a discrete-space $\ell_0$-norm regularizer is introduced to the objective function, which is later approximated with asymptotically tight precision.
The initially non-convex problem is then convexized via fractional programming and solved efficiently via a proximal gradient method.
Simulation results demonstrate that the proposed scheme outperforms both \ac{SotA} techniques, as well as our earlier discrete aware method using a regularizer relaxed by the $\ell_1$-norm.
The contributed scheme can be applied to problems such as recommender systems for streaming or shopping services, as well as social network connections \cite{Chen_2022,Fang_2017,Pech_2017}.

\bibliographystyle{IEEEtran}

\end{document}